\documentclass[a4paper,11pt]{fullverllncs}
\usepackage[left=2.5cm,top=2.5cm,right=2.5cm,centering]{geometry}
\usepackage[dvipdfmx]{graphicx}
\usepackage{amsmath,amssymb}
\usepackage{ascmac}
\usepackage{array}
\usepackage{algpseudocode}
\usepackage{amsfonts}
\usepackage{amssymb}
\usepackage{amsmath}
\usepackage{algorithm, algpseudocode}
\usepackage{multirow}
\usepackage{arydshln}
\usepackage{hhline} 
\usepackage[misc,geometry]{ifsym}

\usepackage[colorlinks]{hyperref, xcolor}
\definecolor{winered}{rgb}{0.5,0,0}
\definecolor{darkblue}{rgb}{0,0,0.5}
\definecolor{darkgreen}{rgb}{0,0.3,0}
\hypersetup{
linkcolor=winered,
citecolor=darkblue,
urlcolor=darkgreen
}
\newcommand{\A}{\mathsf{A}}
\newcommand{\B}{\mathsf{B}}

\newcommand{\C}{\mathsf{C}}

\newcommand{\sfGame}{\mathsf{Game}}

\newcommand{\negl}{\mathsf{negl}}
\newcommand{\poly}{\mathsf{poly}}

\newcommand{\Adv}{\mathsf{Adv}}

\newcommand{\pp}{\mathsf{pp}}
\newcommand{\Z}{\mathbb{Z}}

\newcommand{\G}{\mathbb{G}}
\newcommand{\N}{\mathbb{N}}

\newcommand{\DLIN}{\mathsf{DLIN}}

\newcommand{\sfguess}{\mathsf{guess}}
\newcommand{\sfII}{\mathsf{T\mathchar`-II}}
\newcommand{\sfI}{\mathsf{T\mathchar`-I}}

\newcommand{\GrGen}{\mathsf{GrGen}}
\newcommand{\Ours}{\mathsf{Ours}}

\newcommand{\MsgSp}{\mathsf{MsgSp}}

\newcommand{\OTS}{\mathsf{OTS}}
\newcommand{\OTSKGen}{\mathsf{OTS.KGen}}
\newcommand{\OTSSign}{\mathsf{OTS.Sign}}
\newcommand{\OTSVerify}{\mathsf{OTS.Verify}}
\newcommand{\nOTSKGen}{\mathsf{KGen}}
\newcommand{\nOTSSign}{\mathsf{Sign}}
\newcommand{\nOTSVerify}{\mathsf{Verify}}
\newcommand{\vk}{\mathsf{vk}}
\newcommand{\sk}{\mathsf{sk}}

\newcommand{\sig}{\mathsf{sig}}
\newcommand{\msg}{\mathsf{msg}}

\newcommand{\ek}{\mathsf{ek}}
\newcommand{\dk}{\mathsf{dk}}

\newcommand{\ct}{\mathsf{ct}}
\newcommand{\pt}{\mathsf{pt}}
\newcommand{\sftag}{\mathsf{tag}}
\newcommand{\TagSp}{\mathsf{TagSp}}
\newcommand{\PtSp}{\mathsf{PtSp}}
\newcommand{\CtSp}{\mathsf{CtSp}}
\newcommand{\VkSp}{\mathsf{VkSp}}

\newcommand{\sfINDsTagwCCA}{\mathsf{IND\mathchar`-selTag\mathchar`-wCCA}}
\newcommand{\Dec}{\mathsf{Dec}}

\newcommand{\sOTEUFCMA}{\mathsf{sOT\mathchar`-EUF\mathchar`-CMA}}

\newcommand{\PKEET}{\mathsf{PKEET}}
\newcommand{\nPKEETSetup}{\mathsf{Setup}}
\newcommand{\nPKEETKGen}{\mathsf{KGen}}
\newcommand{\nPKEETEnc}{\mathsf{Enc}}
\newcommand{\nPKEETDec}{\mathsf{Dec}}
\newcommand{\nPKEETTDGen}{{\mathsf{TDGen}}}
\newcommand{\nPKEETTest}{{\mathsf{Test}}}

\newcommand{\PKEETSetup}{\mathsf{PKEET.Setup}}
\newcommand{\PKEETKGen}{\mathsf{PKEET.KGen}}
\newcommand{\PKEETEnc}{\mathsf{PKEET.Enc}}
\newcommand{\PKEETDec}{\mathsf{PKEET.Dec}}
\newcommand{\PKEETTDGen}{{\mathsf{PKEET.TDGen}}}
\newcommand{\PKEETTest}{{\mathsf{PKEET.Test}}}

\newcommand{\sfOWCCATI}{\mathsf{OW\mathchar`-CCA\mathchar`-T\mathchar`-I}}
\newcommand{\sfINDCCATII}{\mathsf{IND\mathchar`-CCA\mathchar`-T\mathchar`-II}}

\newcommand{\td}{\mathsf{td}}

\newcommand{\DDHVerify}{\mathsf{DDHVerify}}

\newcommand{\TBE}{\mathsf{TBE}}
\newcommand{\nTBESetup}{\mathsf{Setup}}
\newcommand{\nTBEEnc}{\mathsf{Enc}}
\newcommand{\nTBEDec}{\mathsf{Dec}}
\newcommand{\TBESetup}{\mathsf{TBE.Setup}}
\newcommand{\TBEEnc}{\mathsf{TBE.Enc}}
\newcommand{\TBEDec}{\mathsf{TBE.Dec}}

\newcommand{\IBE}{\mathsf{IBE}}
\newcommand{\HIBE}{\mathsf{HIBE}}
\newcommand{\IBESetup}{\mathsf{IBE.Setup}}

\newcommand{\IBEEnc}{\mathsf{IBE.Enc}}

\newcommand{\HIBESetup}{\mathsf{HIBE.Setup}}
\newcommand{\HIBEEDer}{\mathsf{HIBE.Der}}
\newcommand{\HIBEEnc}{\mathsf{HIBE.Enc}}
\newcommand{\HIBEDec}{\mathsf{HIBE.Dec}}
\newcommand{\id}{\mathsf{id}}
\newcommand{\mpk}{\mathsf{mpk}}
\newcommand{\msk}{\mathsf{msk}}
\newcommand{\Hash}{\mathsf{Hash}}

\newcommand{\EventForge}{\mathsf{Forge}}
\newcommand{\EventColl}{\mathsf{Coll}}

\newenvironment{proof*}[1]
  {\renewcommand{\proofname}{#1}
   \begin{proof}}
  {\end{proof}}

\spnewtheorem{assumption}{Assumption}{\bfseries}{\itshape}

\begin{document}

\title{Public Key Encryption with Equality Test \\ from Tag-Based Encryption\thanks{A preliminary version of this paper is appeared in the 20th International Workshop on Security (IWSEC 2025).}}
\author{Masayuki Tezuka\textsuperscript{(\Letter)} \and Keisuke Tanaka}
\authorrunning{M.Tezuka et al.}
\titlerunning{Public Key Encryption with Equality Test from Tag-Based Encryption}
\institute{Institute of Science Tokyo, Tokyo, Japan\\
\email{tezuka.m.eab3@m.isct.ac.jp}
}
\maketitle
\pagestyle{plain}
\noindent
\makebox[\linewidth]{September 22, 2025}

\begin{abstract}
Public key encryption with equality test (PKEET), proposed by Yang et al. (CT-RSA 2010), is a variant of public key encryption that enables an equality test to determine whether two ciphertexts correspond to the same plaintext. This test applies not only for ciphertexts generated under the same encryption key but also for those generated under different encryption keys.
To date, several generic constructions of PKEET have been proposed.
However, these generic constructions have the drawback of reliance on the random oracle model or a (hierarchical) identity-based encryption scheme.

In this paper, we propose a generic construction of a PKEET scheme based on tag-based encryption without the random oracle model.
Tag-based encryption is a weaker primitive than identity-based encryption.
Our scheme allows to derive new PKEET schemes without the random oracle model.
By instantiating our construction with the pairing-free tag-based encryption scheme by Kiltz (TCC 2006), we obtain a pairing-free PKEET scheme without the random oracle model.
Moreover, by instantiating our construction with a tag-based encryption scheme based on the learning parity with noise (LPN) assumption, we obtain a PKEET scheme based on the LPN assumption without the random oracle model.

\end{abstract}

\section{Introduction}

\paragraph{\bf Background.}
Public key encryption with equality test (PKEET), proposed by Yang, Tan, Huang, and Wong \cite{YTHW10}, is a variant of public key encryption (PKE) that enables an equality test to determine whether two ciphertexts correspond to the same plaintext. This test applies not only for ciphertexts generated under the same encryption key but also for those generated under different encryption keys.
The original PKEET by Yang et al. \cite{YTHW10} allows anyone to perform an equality test. 
However, this unrestricted capability enables anyone to reveal some information related to the plaintext.\footnote{For example, let us consider the following case. A sender encrypts a plaintext $\pt$ using a public encryption key $\ek$ and generates a ciphertext $\ct$. Then, anyone can choose some plaintext $\pt'$, generate a ciphertext $\ct'$ by using $\ek$, and perform the equality check on $(\ct, \ct')$. 
This reveals the information whether $\pt = \pt'$ or $\pt \neq \pt'$.}
To mitigate this issue, PKEET introduces an equality test restriction that confines equality tests to designated testers.
Due to the usefulness of the equality test property, PKEET has various practical applications, including keyword search on encrypted data, encrypted data partitioning for efficient encrypted data management, personal health record systems, and encrypted databases.

The security of PKEET is defined with respect to two types of adversaries.
A type-I adversary is modeled as a tester who possesses trapdoors, whereas a type-II adversary is modeled as any party other than the sender, receivers, or tester, and thus does not possess trapdoors. 
Since type-I adversaries have trapdoors, indistinguishability security is impossible in this case.
For this reason, previous works have required that PKEET satisfy one-wayness under chosen ciphertext attacks for type I (OW-CCA-T-I) security.
In contrast, for type-II adversaries, PKEET should satisfy indistinguishability under chosen ciphertext attacks for type II (IND-CCA-T-II) security.

\paragraph{\bf Generic Construction of PKEET.}
Several constructions of PKEET schemes and their variants have been proposed, including pairing-based schemes (e.g., \cite{CXTT21,HCT15,HTCLS14,LZL12,MHZY15,MZHY15,QYLZS18,Tan11,YTHW10,ZCLQW19}) and lattice-based schemes (e.g., \cite{DFKRS19,DRSFKS22}).
Lee, Ling, Seo, and Wang \cite{LLSW16} proposed a semi-generic construction based on IND-CCA secure PKE and the computational Diffie-Hellman (CDH) assumption in the random oracle model (ROM) \cite{BR93}.

For generic constructions of PKEET, Lin, Sun, and Qu \cite{LSQ18} proposed a generic construction from IND-CCA secure PKE in the ROM.
Lee, Ling, Seo, and Wang \cite{LLSW19} gave a generic construction from IND-CCA secure PKE with randomness extractability in the ROM.
Lee, Ling, Seo, Wang, and Youn \cite{LLSWY20} proposed a generic construction from hierarchical identity-based encryption (HIBE) without the ROM.
Duong, Roy, Susilo, Fukushima, Kiyomoto, and Sipasseuth~\cite{DRSFKS22} presented the idea of the construction of PKEET from identity-based encryption (IBE) and provided two constructions from specific lattice-based IBE schemes.
However, they only proved the security of PKEET schemes from specific lattice-based IBE schemes and did not provide security proof for the generic construction idea.
Choi, Park, and Lee \cite{CPL25} proposed a generic construction from OW-CPA secure PKE in the ROM.

\subsection{Motivation}
\paragraph{\bf PKEET from Weaker Primitives without the ROM.}
Although some generic PKEET constructions have been proposed, these have weaknesses.
The generic constructions \cite{CPL25,LLSW19,LSQ18} have the drawback of relying on the ROM.
The generic construction of Lee et al. \cite{LLSWY20} relies on $2$-HIBE which is a strong primitive than PKE.
Duong et al. \cite{DRSFKS22} presented the idea of construction of PKEET from the IBE.
This approach may yield a more efficient scheme compared to HIBE-based generic constructions \cite{LLSWY20}. 
However, using IBE does not improve the underlying assumptions, since the existence of HIBE implies the existence of IBE~\cite{DG17}.
To summarize the previous works of generic construction of PKEET \cite{CPL25,DRSFKS22,LLSW19,LLSWY20,LSQ18}, the following question arises:

 \begin{quote}
\emph{Is it possible to construct PKEET from weaker primitives than IBE without the ROM?}
\end{quote}

\subsection{Our Contribution}\label{Sec_OurCon}
\paragraph{\bf PKEET Scheme from Tag-Based Encryption.} 
We propose a generic construction of PKEET from tag-based encryption (TBE) \cite{MRY04} without the ROM.
TBE is a variant of PKE where both the encryption algorithm and the decryption take an additional input called a tag.
The tag in TBE serves a similar role to the identity in IBE. 
However, an important difference is that IBE includes a key derivation algorithm that generates a decryption key corresponding to each identity from a master secret key, whereas TBE lacks such an algorithm and instead uses a single decryption key to decrypt ciphertexts for all tags.
Kiltz \cite{Kil06} pointed out that IBE is unnecessarily strong for obtaining IND-CCA secure PKE, and gave a construction of IND-CCA secure PKE from indistinguishability against selective-tag weak chosen-ciphertext attacks (IND-selTag-wCCA) secure TBE without the ROM.

\paragraph{\bf Comparison with Previous Works.}
We summarize the comparison between our construction $\PKEET_{\Ours}$ and the previous generic constructions \cite{CPL25,DRSFKS22,LLSW19,LLSWY20,LSQ18} in Fig. \ref{Our_result_Comp}.
\begin{figure}[htbp]
\begin{center}
\renewcommand{\arraystretch}{1.2}
\begin{tabular}{clccc}\hline
\begin{tabular}{l}
~~~~~Scheme~~~~~
\end{tabular}
&
\begin{tabular}{l}
Primitives
\end{tabular}
&
\begin{tabular}{c}
ROM\\
\end{tabular}
&
\\

\hline
\begin{tabular}{l}
\cite{LSQ18}
\end{tabular}
&
\begin{tabular}{l}
IND-CCA PKE + Lagrange interporation\\
+ OW\&CR HASH
\end{tabular}
&
\begin{tabular}{l}
\fcolorbox{black}{lightgray}{Yes}\\
\end{tabular}
\\
\hline

\begin{tabular}{l}
\cite{LLSW19}
\end{tabular}
&
\begin{tabular}{l}
IND-CCA PKE with randomness extractability~~~\\
+ OW\&CR HASH
\end{tabular}
&
\begin{tabular}{l}
\fcolorbox{black}{lightgray}{Yes}\\
\end{tabular}
\\

\hline

\begin{tabular}{l}
\cite{LLSWY20}
\end{tabular}
&
\begin{tabular}{l}
\fcolorbox{black}{lightgray}{IND-sID-CPA 2-HIBE}\\
+ OTS + OW\&CR HASH
\end{tabular}
&
\begin{tabular}{c}
No\\
\end{tabular}
\\

\hline
\begin{tabular}{l}
\cite{CPL25}
\end{tabular}
&
\begin{tabular}{l}
OW-CPA PKE\\
+ OW\&CR HASH
\end{tabular}
&
\begin{tabular}{l}
\fcolorbox{black}{lightgray}{Yes}\\
\end{tabular}
\\

\hline
\begin{tabular}{l}
$\PKEET_{\Ours}$\\
Section \ref{SubSecGeneconPKEET}
\end{tabular}
&
\begin{tabular}{l}
IND-selTag-wCCA TBE\\
+ OTS + OW\&CR HASH\\
\end{tabular}
&
\begin{tabular}{c}
No\\
\end{tabular}
\\
\hline
\end{tabular}\\
\caption{\small The comparison result among generic constructions of PKEET schemes.
}
\label{Our_result_Comp}
\end{center}
We highlight weaknesses of the corresponding construction compared to our proposed scheme $\PKEET_{\Ours}$ in \fcolorbox{black}{lightgray}{lightgray}.
In the column ``Primitive'', ``OTS'' (resp., ``OW\&CR HASH'') represents one-time signatures (resp. hash functions with one-wayness and collision resistance properties).
In the column ``ROM'', ``Yes'' represents that the security of the corresponding scheme is proven under the ROM.
 ``No'' represents that the security of the corresponding scheme is proven without the ROM.
\end{figure}
Compared with PKEET constructions in previous works, our construction offers the following strengths:
\begin{itemize}
\item The security of our construction is proven without the ROM. In contrast, the generic constructions in \cite{CPL25,LLSW19,LSQ18} rely on the ROM. 
Our construction requires TBE to satisfy the IND-selTag-wCCA security.
In contrast, the generic construction in \cite{LLSWY20} and idea of construction in \cite{DRSFKS22} rely on strong assumptions such as $2$-HIBE or IBE.
Since the existence IBE of implies the IND-selTag-wCCA secure TBE \cite{Kil06}, our construction is based on weaker assumptions compared to generic constructions in \cite{LLSWY20}.

\item Our generic construction allowed us to derive new PKEET scheme.
For example, Kiltz~\cite{Kil06} gave the pairing-free TBE scheme based on the decision linear (DLIN) assumption on gap groups.
By instantiating our construction with this TBE scheme, we obtain a pairing-free PKEET scheme.
Compared to pairing-based schemes \cite{CXTT21,HCT15,HTCLS14,LZL12,MHZY15,MZHY15,QYLZS18,Tan11,YTHW10,ZCLQW19}, this derived scheme has strength of pairing-free without the ROM.

As an another example, Kiltz, Masny, and Pietrzak \cite{KMP14} proposed TBE scheme based on the learning parity with noise (LPN) assumption.
By instantiating our construction with TBE scheme, we obtain the LPN-based PKEET without the ROM.
We can use another LPN-based TBE by Yu and Zhang \cite{YZ16}.
There is a trade-off between the parameter and efficiency when compared with \cite{KMP14}.
\end{itemize}

\subsection{Technical Overview}\label{Subsec_Tech_Over}

\paragraph{\bf Starting Point: Generic Construction from HIBE.} 
The starting point of our construction is a generic construction of PKEET from $2$-HIBE by Lee et al. \cite{LLSWY20}.
Their construction is derived by applying the CHK transformation \cite{CHK04} to a $2$-HIBE scheme.
Let $\HIBE$ be a $2$-HIBE scheme, $\OTS$ a one-time signature scheme, and $H$ a collision resistance hash function.
Now, we review their PKEET construction. 
To simplify the discussion, we omit the integrity check for a ciphertext in the decryption procedure.

\begin{itemize}
\item A receiver runs $\HIBESetup$ and obtains a tuple of a master public/secret key $(\mpk, \msk)$.
In their PKEET scheme, an encryption key is set as $\ek = \mpk$ and a decryption key as $\dk = \msk$.

\item To encrypt a plaintext $\pt$, the sender first generate a verification/signing key $(\vk, \sk)$ by running $\OTSKGen$.
Then, the the sender generates two ciphertexts $\ct_{0} = \HIBEEnc(\mpk, \id=0.\vk, \allowbreak \pt)$, $\ct_{1} \leftarrow \HIBEEnc(\mpk, \id'=1.\vk, H(\pt))$, and a signature $\sig \leftarrow \OTSSign(\vk, \sk, (\ct_{0}, \ct_{1}))$ where $b.\vk$ represents that the $1$st level identity is $b$ and the $2$nd level of identity is $\vk$ for $b \in \{0, 1\}$. 
The resulting ciphertext is $\ct_{\PKEET} = (\vk, \ct_{0}, \ct_{1}, \sig)$.

\item To encrypt a ciphertext $\ct_{\PKEET} = (\vk, \ct_{0}, \ct_{1}, \sig)$, a receiver decrypts a ciphertext by deriving $\dk_{b.\vk} \leftarrow \HIBEEDer(\msk, b.\vk)$ and decrypting $\pt \leftarrow \HIBEDec(\dk_{0.\vk}, 0.\vk, \ct_{0})$ and $H(\pt) \leftarrow \HIBEDec(\dk_{1.\vk}, \allowbreak 1.\vk, \ct_{1})$.

\item To delegate equality testing to a tester, a receiver derives $\dk_{1} \leftarrow \HIBEEDer(\msk, \allowbreak 1)$ and sends $\dk_{1}$ as a trapdoor to the tester.

\item A tester performs the equality test by decrypting the $\ct_{1}$ by running $\dk_{1.\vk} \leftarrow \HIBEEDer(\msk, 1.\vk)$ and $\pt'_{1} \leftarrow \HIBEDec(\dk_{1.\vk}, 1.\vk, \ct_{1})$. 
By comparing $\pt'_{1}$ values from ciphertexts, the tester determines equality.
\end{itemize}
\paragraph{\bf First Step: Generic Construction from IBE.} 
In the general construction from a $2$-leveled HIBE scheme, only $0$ and $1$ are used for $1$st leveled identity (HIBE prefixes).
It does not appear to make use of the full power of $2$-leveled HIBE.
From this fact, with a simple modification to the construction \cite{LLSWY20}, we obtain a generic construction of a PKEET scheme from an IBE scheme.
Let $\IBE$ and  $\IBE'$ be IBE schemes.
Instead of encrypting $\pt$ and $H(\pt)$ with different HIBE prefixes, we prepare two master public keys $(\mpk, \mpk')$ by running the setup algorithms of two IBE schemes and encrypt $\pt$ and $H(\pt)$ with separate IBE maser public key $\mpk$ and  $\mpk'$, respectively.
We present this construction as follows.

\begin{itemize}
\item A receiver runs $\IBESetup$, $\IBESetup'$ in parallel and obtains two pairs of a master public/secret key $(\mpk_{\IBE}, \msk_{\IBE})$, $(\mpk_{\IBE'}, \msk_{\IBE'})$. 
An encryption key is set as  $\ek = (\mpk_{\IBE}, \allowbreak \mpk_{\IBE'})$ and a decryption key as $\dk = (\msk_{\IBE}, \msk_{\IBE'})$.

\item To encrypt a plaintext $\pt$, the sender generates a verification/signing key $(\vk, \sk)$ by running $\OTSKGen$, two ciphertexts $\ct = \IBEEnc(\mpk_{\IBE}, \id =\vk, \allowbreak \pt)$, $\ct' = \IBEEnc'(\mpk_{\IBE'}, \id =\vk, H(\pt))$.
The resulting ciphertext is $\ct_{\PKEET} = (\vk, \ct, \ct', \sig)$ where $\sig = \OTSSign(\vk, \sk, \allowbreak (\ct, \ct'))$.

\item A receiver decrypt a ciphertext by using $\dk = (\msk_{\IBE}, \msk_{\IBE'})$.

\item To delegate equality testing to a tester, a receiver sends $\td = \msk_{\IBE'}$ as a trapdoor.

\item A tester performs the equality test by decrypting the $\ct'$ part of $\ct_{\PKEET}$ by using $\msk_{\IBE'}$. 
By comparing resulting $\pt'$, the tester determines equality.
\end{itemize}
\paragraph{\bf Second Step: Replacing IBE with TBE.} 
Since Kiltz~\cite{Kil06} achieved a CCA secure PKE from a TBE via the CHK transformation, this fact suggests that the above construction could be modified to use a TBE instead of an IBE.
Let $\TBE$ and  $\TBE'$ be TBE schemes.
Now, we present the overview of our construction.

\begin{itemize}
\item A receiver runs $\TBESetup$, $\TBESetup'$ in parallel and obtains two tuples of an encryption/decryption key $(\ek_{\TBE}, \dk_{\TBE})$, $(\ek'_{\TBE}, \dk'_{\TBE'})$. 
An encryption key is set as  $\ek = (\ek_{\TBE}, \allowbreak \ek'_{\TBE'})$ and a decryption key as $\dk = (\dk_{\TBE}, \dk'_{\TBE'})$.

\item To encrypt a plaintext $\pt$, the sender generates a verification/signing key $(\vk, \sk)$ by running $\OTSKGen$, two ciphertexts $\ct = \TBEEnc(\ek_{\TBE}, \allowbreak \sftag =\vk, \allowbreak \pt)$, $\ct' = \TBEEnc'(\mpk'_{\TBE'}, \sftag =\vk, H(\pt))$.
The resulting ciphertext is $\ct_{\PKEET} = (\vk, \ct, \ct', \sig)$ where $\sig = \OTSSign(\vk, \sk, \allowbreak (\ct, \ct'))$.

\item A receiver decrypt a ciphertext by using $(\dk_{\TBE}, \dk_{\TBE'})$.

\item To delegate equality testing to a tester, the receiver sends $\td = \dk_{\TBE'}$ as a trapdoor.

\item The tester performs the equality test by decrypting the $\ct'$ part of $\ct_{\PKEET}$ by using $\dk_{\TBE'}$. 
By comparing resulting $\pt'$, the tester determines equality.
\end{itemize}

\subsection{RoadMap}
In Section \ref{SecPrelimi},  we introduce notations and review the definition of a TBE scheme and its security notions.
In Section \ref{SecPKEETDef},  we review the definition of a PKEET scheme and its security notions.
In Section \ref{SecOurPKEETConst}, we propose a generic construction of PKEET from TBE.
Then, we prove the security of our construction.
In Appendix \ref{Sec:Fundamental_Primitives}, we review fundamental security notions of hash function and one-time signature.
In Appendix \ref{Sec:DLIN_TBE}, we review the DLIN-based TBE scheme by Kiltz \cite{Kil06}.

\section{Preliminaries}\label{SecPrelimi}
In this section, we introduce notations and review the definition of a tag-based encryption (TBE) scheme and its security notions.
Please refer to Appendix \ref{Sec:Fundamental_Primitives} for the fundamental of hash functions and one-time signatures.

\subsection{Notations}
Let $\lambda$ be the security parameter. 
A function $f(\lambda)$ is negligible in $\lambda$ if $f(\lambda)$ tends to $0$ faster than $\frac{1}{\lambda^c}$ for every constant $c > 0$ (i.e., $f(\lambda) = \lambda^{-\omega(1)}$).
Let $\negl(\lambda)$ denote a negligible function in $\lambda$ and $\poly(\lambda)$ denote a polynomial function in $\lambda$.
For a finite set $S$, $s \xleftarrow{\$}S$ represents that an element $s$ is chosen from $S$ uniformly at random, and $|S|$ represents the number of elements in $S$.
For an algorithm $\A$, $y \leftarrow \A(x)$ denotes that the algorithm $\A$ outputs $y$ on input~$x$.
When we explicitly show that $\A$ uses randomness $r$, we write $y \leftarrow \A(x; r)$.
We abbreviate probabilistic polynomial time as PPT.

We use standard code-based security games.
A game $\sfGame$ is a probability experiment between a challenger $\C$ and an adversary $\A$.
We denote the output $b \in \{0, 1\}$ of game $\sfGame$ for $\A$ as $\sfGame_{\A} \Rightarrow b$.
We say that $\A$ wins the game $\sfGame$ if $\sfGame_{\A} \Rightarrow 1$.

\subsection{Tag-Based Encryption}
We review a definition of a tag-based encryption scheme and its security notion~\cite{Kil06}.
\begin{definition}[Tag-Based Encryption Scheme \cite{Kil06}]\label{Def_IBE}
A tag-based encryption scheme $\TBE$ with an identity space $\TagSp(\lambda)$ and a plaintext space $\PtSp(\lambda)$ is a tuple of PPT algorithms $(\nTBESetup, \nTBEEnc, \nTBEDec)$.
\begin{itemize}
\item $\nTBESetup(1^{\lambda}):$ A setup algorithm (probabilistic) takes as an input a security parameter $1^{\lambda}$.
It returns an encryption key and a decryption key $(\ek, \dk)$.
\item $\nTBEEnc(\ek, \sftag, \pt):$ An encryption algorithm (probabilistic) takes as an encryption key $\ek$, a tag $\sftag$, and a plaintext $\pt$.
It returns a ciphertext~$\ct$.
\item $\nTBEDec(\dk, \sftag, \ct):$ A decryption algorithm (probabilistic) takes as an input a decryption key $\dk$, a tag $\sftag$ and a ciphertext $\ct$.
It returns a plaintext $\pt$ or~$\bot$.
\end{itemize}
\end{definition}

For $\TBE$, we require the following correctness.

\paragraph{\bf Correctness:}
$\TBE$ has correctness, if for all $\lambda \in \N$, for all $\sftag \in \TagSp(\lambda)$, $\pt \in \PtSp(\lambda)$, $(\ek, \dk) \leftarrow \nTBESetup(1^{\lambda})$, and $\ct \leftarrow \nTBEEnc(\ek, \sftag, \pt)$, $\nTBEDec(\dk, \sftag, \ct) = \pt$ holds.

\begin{definition}[IND-sTag-wCCA Security \cite{Kil06}]\label{Def_IND_sID_CPA_PKEET}
Let $\TBE$ be a tag-based encryption scheme and $\A$ be a PPT algorithm.
Indistinguishability against selective-tag weak chosen-ciphertext attacks (IND-selTag-wCCA) security is defined by the following IND-sTag-wCCA game $\sfGame^{\sfINDsTagwCCA}_{\TBE, \A}(1^{\lambda})$ between a challenger $\C$ and an adversary $\A$.
\begin{itemize}
\item $\A$ sends a target tag $\sftag^{*} \in \TagSp(\lambda)$ to $\C$.
\item $\C$ runs $(\ek, \dk) \leftarrow \nTBESetup(1^{\lambda})$ and sends $\ek$ to $\A$.
\item $\A$ makes queries for the oracle $\mathcal{O}^{\Dec}$ polynomially many times.
\begin{itemize}
\item For a decryption query on $(\sftag, \ct)$, if $\sftag \neq \sftag^{*}$ holds, $\mathcal{O}^{\Dec}$ returns $\pt \leftarrow  \nTBEDec(\dk, \sftag, \ct)$.
Otherwise, it returns $\bot$.
\end{itemize}
\item $\A$ sends a challenge $(\pt_{0}, \pt_{1})$ to $\C$.
\item $\C$ samples $b^{*} \xleftarrow{\$} \{0, 1\}$ and returns $\ct^{*}_{b^*} \leftarrow \nTBEEnc(\ek, \sftag^{*}, \pt_{b^{*}})$ to $\A$.
\item $\A$ makes queries for the oracle $\mathcal{O}^{\Dec}$ polynomially many times. 
\item $\A$ finally outputs a guess $b^{*}_{\sfguess}$ to $\C$.
\item If $b^{*}_{\sfguess} = b^{*}$, $\A$ wins the game.
\end{itemize}

The advantage of an adversary $\A$ for the IND-selTag-wCCA security game is defined by $\Adv^{\sfINDsTagwCCA}_{\TBE, \A}(\lambda):= \Pr[\sfGame^{\sfINDsTagwCCA}_{\TBE, \A}(1^{\lambda}) \Rightarrow 1]$.
$\TBE$ satisfies the IND-selTag-wCCA security if for any PPT adversary $\A$, $\Adv^{\sfINDsTagwCCA}_{\TBE, \A}(\lambda)$ is $\negl(\lambda)$.
\end{definition}

\section{Public Key Encryption with Equality Test}\label{SecPKEETDef}
In this section, we review a definition of a public key encryption scheme with equality test (PKEET) and its security notions.

\subsection{Definition of PKEET}
We review a definition of a PKEET scheme.
\begin{definition}
A public key encryption scheme with equality test $\PKEET$ with a plaintext space $\PtSp(\lambda)$ is a tuple of PPT algorithms $(\nPKEETSetup, \nPKEETKGen, \nPKEETEnc, \allowbreak \nPKEETDec, \nPKEETTDGen, \allowbreak \nPKEETTest)$.
\begin{itemize}
\item $\nPKEETSetup (1^{\lambda}):$ A setup algorithm takes as an input a security parameter $1^{\lambda}$. 
It returns a public parameter $\pp$.
We implicitly assume that the algorithms described below take a public parameter $\pp$ as input and omit $\pp$ from the input of all algorithms except for $\nPKEETKGen$.
\item $\nPKEETKGen (\pp):$ A key-generation algorithm takes as an input a public parameter $\pp$. 
It returns an encryption key $\ek$ and a decryption key $\dk$.
\item $\nPKEETEnc (\ek, \pt):$ An encryption algorithm takes as an input an encryption key $\ek$ and a plaintext $\pt$. 
It returns a ciphertext $\ct$.
\item $\nPKEETDec(\dk, \ct):$ A decryption algorithm takes as an input a decryption key $ \dk$ and a ciphertext $\ct$.
It returns a plaintext $\pt$ or $\bot$.
\item $\nPKEETTDGen(\dk):$ A trapdoor generation algorithm takes as an input a decryption key $\dk$.
It returns a trapdoor $\td$.
\item $\nPKEETTest:((\ek^{(\theta)}, \ct^{(\theta)}, \td^{(\theta)})_{\theta \in \{0, 1\}}):$ A test algorithm takes as an input tuples of $(\ek^{(\theta)}, \ct^{(\theta)}, \td^{(\theta)})_{\theta \in \{0, 1\}}$.
It output a bit $b \in \{0, 1\}$.
\end{itemize}
\end{definition}

For $\PKEET$, we require the following correctness.

\paragraph{\bf Decryption correctness:}
$\PKEET$ satisfies correctness for decryption, if for all $\lambda \in \mathbb{N}$, $\pt \in \PtSp(\lambda)$, $\pp \leftarrow \nPKEETSetup (1^{\lambda})$, $(\ek, \allowbreak \dk) \leftarrow \nPKEETKGen(\pp)$, and $\ct \leftarrow \nPKEETEnc (\ek, \pt)$, $\nPKEETDec(\dk, \ct) = \pt$ holds.

\paragraph{\bf Equality test correctness for the same plaintext:}
$\PKEET$ satisfies equality test correctness error for the same plaintext, if for all $\lambda \in \N$, for all $\pt \in \PtSp(\lambda)$, $\pp \leftarrow \nPKEETSetup (1^{\lambda})$, $(\ek^{(\theta)}, \dk^{(\theta)}) \leftarrow \nPKEETKGen(\pp)$, $\ct^{(\theta)} \leftarrow  \nPKEETEnc (\ek^{(\theta)}, \allowbreak \pt)$, and $\td^{(\theta)} \leftarrow\nPKEETTDGen(\dk^{(\theta)})$ for $\theta \in \{0, 1\}$, $\nPKEETTest ( (\ek^{(\theta)}, \ct^{(\theta)}, \td^{(\theta)})_{\theta \in \{0, 1\}})= 1$ holds.

\paragraph{\bf Equality test correctness error for different plaintexts:}
$\PKEET$ has equality test correctness error $\delta(\lambda)$ for different plaintexts, if for all $\lambda \in \N$, for all $\pt^{(0)}, \pt^{(1)} \in \PtSp(\lambda)$ such that $\pt^{(0)} \neq \pt^{(1)}$, 
\begin{equation*}
\Pr \left[
\nPKEETTest\left( \left(
\begin{split}
&\ek^{(\theta)}\\
&\ct^{(\theta)}\\
&\td^{(\theta)}
\end{split}
\right)_{\theta \in \{0, 1\}}
\right)
= 1
:
\begin{split}
&\pp \leftarrow \nPKEETSetup (1^{\lambda}), \\
&((\ek^{(\theta)}, \allowbreak \dk^{(\theta)}) \leftarrow \nPKEETKGen(\pp),\\
&~~\ct^{(\theta)} \leftarrow \nPKEETEnc (\ek^{(\theta)}, \pt^{(\theta)}),\\
&~~\td^{(\theta)} \allowbreak \leftarrow\nPKEETTDGen(\dk^{(\theta)}))_{\theta \in \{0, 1\}}
\end{split}
\right] \leq \delta(\lambda)
\end{equation*}
holds.
The probability is taken with respect to the randomness of $\nPKEETSetup$, $\nPKEETKGen$, $\nPKEETEnc$, $\nPKEETTDGen$, $\nPKEETDec$, and $\nPKEETTest$.
We say that $\PKEET$ satisfies the correctness for different plaintexts, if the correctness error $\delta(\lambda)$ is $\negl(\lambda)$.

\subsection{Security of PKEET}
For the security of PKEET, we should consider the following two types of adversaries.
The type-I adversary is modeled as a tester that has a trapdoor for the target receiver.
For these adversaries, it is impossible to satisfy the indistinguishability security for this type of adversary.
Instead, we require a PKEET scheme to satisfy the OW-CCA security for type-I adversaries.

\begin{definition}[OW-CCA Type-I Adversary]\label{Def_OW_CCA_PKEET}
Let $\PKEET$ be a public key encryption scheme with equality test and $\A$ be a PPT algorithm.
One-wayness under chosen ciphertext attacks (OW-CCA) security for a type-I adversary is defined by the following OW-CCA-T-I game $\sfGame^{\sfOWCCATI}_{\PKEET, \A}(1^{\lambda})$ between the challenger $\C$ and an adversary $\A$.
\begin{itemize}
\item $\C$ runs $\pp \leftarrow \nPKEETSetup (1^{\lambda})$, $(\ek, \dk) \leftarrow \nPKEETKGen(\pp)$, $\td \leftarrow \nPKEETTDGen(\dk)$, and sends $(\pp, \ek, \td)$ to $\A$.
\item $\A$ makes queries for the following oracle $\mathcal{O}^{\Dec}$ polynomially many times.
\begin{itemize}
\item For a decryption query on $\ct$ from $\A$, the decryption oracle $\mathcal{O}^{\Dec}$ returns $\pt \leftarrow \nPKEETDec(\dk, \ct)$ to $\A$.
\end{itemize}
\item $\A$ sends an instruction $\mathtt{challenge}$ to $\C$.
\item $\C$ samples $\pt^{*} \xleftarrow{\$} \MsgSp(\lambda)$ and returns $\ct^{*} \leftarrow \nPKEETEnc (\ek,\pt^{*})$ to $\A$.
\item $\A$ makes queries for the following oracle $\mathcal{O}^{\Dec^{*}}$ polynomially many times. 
\begin{itemize}
\item For a decryption query on $\ct$ from $\A$, if $\ct \neq \ct^{*}$ holds, $\mathcal{O}^{\Dec^{*}}$ returns $\pt \leftarrow \nPKEETDec(\dk, \ct)$ to $\A$.
Otherwise, it returns $\bot$.
\end{itemize}
\item $\A$ finally outputs a plaintext $\pt^{*}_{\sfguess}$ to $\C$.
\item If $\pt^{*}_{\sfguess} = \pt^{*}$, $\A$ wins the game.
\end{itemize}
The advantage of an adversary $\A$ for the OW-CCA-T-I  security game is defined by $\Adv^{\sfOWCCATI}_{\PKEET, \A}(\lambda) \allowbreak := \left|\Pr[\sfGame^{\sfOWCCATI}_{\PKEET, \A}(1^{\lambda}) \Rightarrow 1]  \right|$.
$\PKEET$ satisfies the OW-CCA-T-I security if for any PPT adversary $\A$, $\Adv^{\sfOWCCATI}_{\PKEET, \A}(\lambda)$ is $\negl(\lambda)$.
\end{definition}
We note that it is impossible for $\PKEET$ scheme with $|\PtSp^{\PKEET}| = \poly(\lambda)$ to satisfy the OW-CCA security for type-I adversaries.\footnote{For example, let us consider an algorithm that encrypts all message in $\PtSp^{\PKEET}$ and performs an equality test with the target ciphertext $\ct^{*}$.
In the case of $|\PtSp^{\PKEET}| = \poly(\lambda)$, the running time of this algorithm is polynomial time.
This algorithm breaks the OW-CCA security for type-I adversaries security with probability~$1$.}
For this reason, we require $|\PtSp^{\PKEET}| = 2^{\Omega(\lambda)}$.

The type-II adversary is modeled as an adversary who does not have a trapdoor for the target receiver.
We require the IND-CCA security for type-II adversaries for a PKEET scheme.
Now, we define the IND-CCA Type-II security.

\begin{definition}[IND-CCA Type-II Adversary]\label{Def_IND_CCA_PKEET}
Let $\PKEET$ be a public key encryption scheme with equality test and $\A$ be a PPT algorithm.
Indistinguishability under chosen ciphertext attacks (IND-CCA) security for a type-II adversary is defined by the following IND-CCA-T-II game $\sfGame^{\sfINDCCATII}_{\PKEET, \A}(1^{\lambda})$ between the challenger $\C$ and an adversary $\A$.
\begin{itemize}
\item $\C$ runs $\pp \leftarrow \nPKEETSetup (1^{\lambda})$, $(\ek, \dk) \leftarrow \nPKEETKGen(\pp)$, and sends $(\pp, \ek)$ to $\A$.
\item $\A$ makes queries for the following oracle $\mathcal{O}^{\Dec}$ polynomially many times. $\mathcal{O}^{\Dec}$ is the same as described in Definition~\ref{Def_OW_CCA_PKEET}.
\item $\A$ sends a challenge $(\pt_{0}, \pt_{1}) \in \PtSp(\lambda) \times \PtSp(\lambda)$ to $\C$.
\item $\C$ samples $b^{*} \xleftarrow{\$} \{0, 1\}$ and returns $\ct^{*}_{b^*} \leftarrow \nPKEETEnc (\ek,\pt_{b^{*}})$ to $\A$.
\item $\A$ makes queries for the following oracle $\mathcal{O}^{\Dec^{*}}$ polynomially many times. 
$\mathcal{O}^{\Dec^{*}}$ is the same as described in Definition~\ref{Def_OW_CCA_PKEET}.
\item $\A$ finally outputs a guess $b^{*}_{\sfguess}$ to $\C$.
\item If $b^{*}_{\sfguess} = b^{*}$, $\A$ wins the game.
\end{itemize}
The advantage of an adversary $\A$ for the IND-CCA-T-II  security game is defined by $\Adv^{\sfINDCCATII}_{\PKEET, \A}(\lambda) \allowbreak := |\Pr[\sfGame^{\sfINDCCATII}_{\PKEET, \A}(1^{\lambda}) \Rightarrow 1]  - \frac{1}{2} |$.
$\PKEET$ satisfies the IND-CCA-T-II security if for any PPT adversary $\A$, $\Adv^{\sfINDCCATII}_{\PKEET, \A}(\lambda)$ is $\negl(\lambda)$.
\end{definition}

\section{Our Construction of PKEET}\label{SecOurPKEETConst}
First, we give a generic construction of PKEET scheme $\PKEET_{\Ours}$.
Then, we prove the security of our scheme.
\subsection{Generic Construction of PKEET Scheme}\label{SubSecGeneconPKEET}
Our generic construction $\PKEET_{\Ours}$ with a plaintext space $\PtSp^{\PKEET}$ is obtained from the following primitives and the parameters setting.
\begin{itemize}
\item Tag-based encryption schemes $\TBE = (\TBESetup, \TBEEnc, \allowbreak \TBEDec)$ (resp. $\TBE' = (\TBESetup', \allowbreak \TBEEnc', \allowbreak \TBEDec')$) with a tag space $\TagSp_{\TBE}$ (resp. $\TagSp_{\TBE'}$), a plaintext space $\PtSp_{\TBE}$ (resp. $\PtSp_{\TBE'}$), and a ciphertext space $\CtSp_{\TBE}$ (resp. $\CtSp_{\TBE'}$).
\item A (one-time) signature scheme $\OTS=(\nOTSKGen, \allowbreak \nOTSSign, \nOTSVerify)$ with a message space $\MsgSp_{\OTS}$ and a verification key space $\VkSp_{\OTS}$.
\item A family of hash functions $\mathcal{H}= \{H_{i \in I}: X_{\Hash} \rightarrow Y_{\Hash} \}$.
\item $\PtSp_{\PKEET} = \PtSp_{{\TBE}}$ with $|\PtSp_{\PKEET}| = 2^{\Omega(\lambda)}$ for any constant $c>0$, $\TagSp_{\TBE} = \TagSp_{\TBE'} = \VkSp_{\OTS}$, $\PtSp_{\TBE'} =  Y_{\Hash}$, $\MsgSp_{\OTS} =  \CtSp_{{\TBE}} \times \CtSp_{\TBE'}$.
\end{itemize}

Our PKEET schemes $\PKEET_{\Ours}[\TBE, \TBE', \OTS, \mathcal{H}]$ is given in Fig. \ref{PKEET_OurConst}.

\begin{figure}[h]
\centering
\renewcommand{\arraystretch}{1.2}
\begin{tabular}{|l|}
\hline
$\PKEETSetup (1^{\lambda}):$\\
~~$H \xleftarrow{\$} \mathcal{H}$. 
$\pp \leftarrow H$, return $\pp$.\\

$\PKEETKGen(\pp):$\\
~~$(\ek_{\TBE}, \dk_{\TBE}) \leftarrow  \TBESetup(1^{\lambda})$, $(\ek_{\TBE'}, \dk_{\TBE'}) \leftarrow  \TBESetup'(1^{\lambda})$.\\
~~Return $(\ek, \dk) = ((\ek_{\TBE}, \ek_{\TBE'}), (\dk_{\TBE}, \dk_{\TBE'}))$.\\

$\PKEETEnc(\ek=(\ek_{\TBE}, \ek_{\TBE'}), \pt):$\\
~~$(\vk, \sk) \leftarrow \OTSKGen(1^{\lambda})$, $\pt' \leftarrow H(\pt)$, $\ct \leftarrow \TBEEnc(\ek, \vk, \pt)$,\\
~~$\ct' \leftarrow \TBEEnc'(\ek_{\TBE'}, \vk, \pt')$, $\sig \leftarrow \OTSSign(\sk, (\ct, \ct'))$.\\
~~Return $\ct_{\PKEET} \leftarrow (\vk, \ct, \ct', \sig)$.\\

$\PKEETDec(\dk = (\dk_{\TBE}, \dk_{\TBE'}),  \ct_{\PKEET} = (\vk, \ct, \ct', \sig)):$\\
~~If $\OTSVerify(\vk, (\ct, \ct'), \sig) = 0$ return $\bot$.\\
~~$\pt \leftarrow \TBEDec(\dk_{\TBE}, \vk, \ct)$, $\pt' \leftarrow \TBEDec'(\dk_{\TBE'}, \vk, \ct')$.\\
~~If $\pt' = H(\pt)$, return $\pt$. Otherwise return $\bot$.\\

$\PKEETTDGen(\dk =(\dk_{\TBE}, \dk_{\TBE'})):$\\
~~Return $\td = \dk_{\TBE'}$.\\

$\PKEETTest((\ek^{(\theta)} = (\ek^{(\theta)}_{\TBE}, \ek^{(\theta)}_{\TBE'}), $\\
~~~~~~~~~~~~~~~~~~~~~~~~~$\ct_{\PKEET}^{(\theta)}=(\vk^{(\theta)}, \ct^{(\theta)}, \ct'^{(\theta)}, \sig^{(\theta)}), \td^{(\theta)} =  \dk^{(\theta)}_{\TBE'})_{\theta \in \{0, 1\}}):$\\
~~For $\theta \in \{0, 1 \}$, $\pt'^{(\theta)} \leftarrow \TBEDec'(\dk^{(\theta)}_{\TBE'}, \vk, \ct'^{(\theta)})$.\\
~~If $\pt'^{(0)} = \pt'^{(1)}$, return $1$. Otherwise return $0$.\\
\hline
\end{tabular}
\caption{\small Our PKEET scheme construction $\PKEET_{\Ours}[\TBE, \TBE', \OTS, \mathcal{H}]$.}
\label{PKEET_OurConst}
\end{figure}

\paragraph{\bf Correctness:}
Clearly, the correctness of decryption is followed by the correctness of $\TBE$, $\TBE'$, and $\OTS$.
It is also clear that the correctness of equality test for the same plaintext is followed by the correctness of $\TBE'$ and $\OTS$.
The correctness of equality test for different plaintexts is followed by the collision resistance property of $\mathcal{H}$.

\paragraph{\bf Pairing-Free Group Instantiation:}
Our generic construction allows us to derive a pairing-free instantiation.
For example, Kiltz~\cite{Kil06} gave a pairing-free IND-selTag-wCCA secure TBE scheme based on the decision linear (DLIN) assumption on gap groups without the ROM. 
By instantiating our construction with this TBE scheme, we obtain a pairing-free PKEET scheme.
Please refer to Appendix \ref{Sec:DLIN_TBE} or \cite{Kil06} for their DLIN-based TBE construction.

\paragraph{\bf LPN-Based Instantiation:}
Kiltz, Masny, and Pietrzak \cite{KMP14} proposed an IND-selTag-wCCA secure TBE scheme based on the learning parity with noise (LPN) assumption.
This scheme has a tag space $\TagSp = \mathbb{F}_{2^{n}} \backslash \{0\}$ and a message space $\MsgSp = \mathbb{Z}^{n}_{2}$ with $n = \Theta(\lambda^{2})$ where $\lambda$ is the security parameter.
Thus, requirement of $\TagSp = 2^{\Omega(\lambda)}$ is satisfiable.
By instantiating our construction with this TBE scheme, we obtain the LPN-based PKEET.
Please refer to \cite{KMP14} for their LPN-based TBE construction.

\subsection{OW-CCA Security Analysis}
We prove that $\PKEET_{\Ours}$ satisfies OW-CCA-T-I security.

\begin{theorem}[OW-CCA Security Type-I Adversary]\label{Theorem_OW_CCA_Our}
Let $\PKEET_{\Ours}$ be a PKEET scheme with a plaintext space $\PtSp^{\PKEET}$ where $|\PtSp^{\PKEET}|$ is super-polynomial (i.e. $|\PtSp^{\PKEET}| = 2^{\Omega(\lambda)}$ for any constant $c > 0$).
If $\TBE$ satisfies the IND-selTag-wCCA security, $\OTS$ satisfies the sOT-EUF-CMA security, and $\mathcal{H}$ is a family of one-way functions with the collision resistance property, then $\PKEET_{\Ours}$ satisfies the OW-CCA security for type-I adversaries.
\end{theorem}

\begin{proof*}{Proof of Theorem \ref{Theorem_OW_CCA_Our}}
Let $\A$ be an adversary for the OW-CCA-T-1 security for $\PKEET_{\Ours}$ and $\C^{\PKEET}$ be the challenger of OW-CCA-T-I security game.
We prove the OW-CCA-T-I security for $\PKEET_{\Ours}$ by considering the following sequential of games $(\sfGame^{\sfI}_{i, \A})_{i \in \{0, \dots, 3\}}$.
\begin{itemize}
\item{$\sfGame^{\sfI}_{0, \A}:$} 
The original OW-CCA-T-I game $\sfGame^{\sfOWCCATI}_{\PKEET_{\Ours}, \A}(1^{\lambda})$.

\item{$\sfGame^{\sfI}_{1, \A}:$} 
This game is identical to $\sfGame^{\sfI}_{0, \A}$ except that we change the timing of generating of $(\vk^{*}, \sk^{*})$. This tuple is used to generate a challenge ciphertext $\ct_{\PKEET}^{*} = (\vk^{*}, \ct^{*}, \ct'^{*}, \sig)$.
Moreover, we introduce the event $\EventForge$ in $\sfGame^{\sfI}_{0, \A}$.

At the beginning of the game, $\C^{\PKEET}$ runs $(\vk^{*}, \sk^{*}) \leftarrow \OTSKGen(1^{\lambda})$.
Let $\EventForge$ be the event that $\A$ makes a decryption query that satisfies either of the two following conditions.
\begin{itemize}
\item Before the challenge query, $\A$ makes a decryption query on a ciphertext $\ct_{\PKEET}=(\vk, \ct, \ct', \sig)$ such that $\vk = \vk^{*}$ and $\OTSVerify(\vk^{*}, (\ct, \ct'), \allowbreak \sig) =1$ holds.
\item After the challenge query, $\A$ makes a decryption query on a ciphertext $\ct_{\PKEET}=(\vk, \ct, \ct', \sig)$
such that $\vk = \vk^{*} \land \ct_{\PKEET} \neq \ct_{\PKEET}^{*} \land \OTSVerify(\vk^{*}, \allowbreak(\ct, \allowbreak \ct'),\allowbreak \sig) =1$.
\end{itemize}

\item{$\sfGame^{\sfI}_{2, \A}:$} 
This game is identical to $\sfGame^{\sfI}_{1, \A}$ except that we change the generation of the challenge ciphertext $\ct_{\PKEET}^{*} = (\vk^{*}, \ct^{*}, \ct'^{*}, \sig)$ and the winning condition of $\A$.
For a challenge query $\mathtt{challenge}$ from $\A$, $\C^{\PKEET}$ samples $\pt^{*} \xleftarrow{\$} \PtSp^{\PKEET}(\lambda)$, sets $\pt'^{*} \leftarrow H(\pt^{*})$, $\ct^{*} \leftarrow \TBEEnc(\ek_{\TBE}, \allowbreak \vk^{*}, \allowbreak 0)$, $\ct'^{*} \leftarrow \TBEEnc'(\ek_{\TBE'}, \allowbreak \vk^{*}, \allowbreak \pt'^{*})$.
For the final output $\pt^{*}_{\sfguess}$ by $\A$, $\A$ wins the $\sfGame^{\sfI}_{2, \A}$ if $\pt_{\sfguess}^{*} = \pt^{*}$ holds. 

\item{$\sfGame^{\sfI}_{3, \A}:$} 
This game is identical to $\sfGame^{\sfI}_{2, \A}$ except that we change the winning condition for $\A$.
For the final output $\pt^{*}_{\sfguess}$ by $\A$, if $H(\pt_{\sfguess}^{*}) = H(\pt^{*})$ holds, $\A$ wins the $\sfGame^{\sfI}_{3, \A}$. 
\end{itemize}

For $i \in \{0, \dots, 3\}$, let $\sfGame^{\sfI}_{i, \A} \Rightarrow 1$ be the event that $\sfGame^{\sfI}_{i, \A}$ outputs $1$.
Then, the following lemmas hold.
\begin{lemma}\label{OWG0toG1}
If $\OTS$ satisfies the sOT-EUF-CMA security, \\
\begin{equation*}
|\Pr[\sfGame^{\sfI}_{1, \A} \Rightarrow 1 ] - \allowbreak \Pr[\sfGame^{\sfI}_{0, \A} \Rightarrow 1] | = \negl(\lambda)
\end{equation*} 
holds.
\end{lemma}

\begin{proof*}{Proof of Lemma \ref{OWG0toG1}}
We confirm that Lemma \ref{OWG0toG1} holds.
Under the condition where the event $\EventForge$ does not occurs, $\sfGame^{\sfI}_{0, \A}$ and $\sfGame^{\sfI}_{1, \A}$ are identical.
We see that $\Pr[\sfGame^{\sfI}_{0, \A}\Rightarrow 1|\lnot\EventForge] = \Pr[\sfGame^{\sfI}_{1, \A}\Rightarrow 1|\lnot\EventForge]$ holds.
Then, we have 
\begin{equation*}
\begin{split}
|\Pr[\sfGame^{\sfI}_{1, \A}\Rightarrow 1] - \Pr[\sfGame^{\sfI}_{0, \A}\Rightarrow 1]| \leq  \Pr[\EventForge].
\end{split}
\end{equation*}

Since $\OTS$ satisfies the sOT-EUF-CMA security, $\Pr[\EventForge] = \negl(\lambda)$ holds.
\par \hfill (Lemma \ref{OWG0toG1}) \qed
\end{proof*}

\begin{lemma}\label{OWG1toG2}
$\TBE$ satisfies the IND-selTag-wCCA security, 
\begin{equation*}
|\Pr[\sfGame^{\sfI}_{2, \A} \allowbreak \Rightarrow 1 ]  - \allowbreak \Pr[\sfGame^{\sfI}_{1, \A} \Rightarrow 1]| = \negl(\lambda)
\end{equation*}
holds.
\end{lemma}

\begin{proof*}{Proof of Lemma \ref{OWG1toG2}}
We prove Lemma \ref{OWG1toG2} by constructing the following reduction algorithm $\B$.
Let $\C^{\TBE}$ the challenger of the IND-selTag-wCCA security game of $\TBE$.
We briefly explain how to construct $\B$ as follows.

\begin{itemize}
\item Given a security parameter $1^{\lambda}$, $\B$ runs $(\vk^{*}, \sk^{*}) \leftarrow \OTSKGen(1^{\lambda})$ and sends $\vk^{*}$ to $\C^{\TBE}$. Then, $\B$ receives $\ek^{*}_{\TBE}$ from $\C^{\TBE}$.

\item $\B$ samples $H \xleftarrow{\$} \mathcal{H}$, $\pp \leftarrow H$, $(\ek_{\TBE'}, \dk_{\TBE'}) \allowbreak \leftarrow  \nTBESetup'(1^{\lambda})$, $\ek \leftarrow (\ek^{*}_{\TBE}, \ek_{\TBE'})$, $\td \leftarrow \dk_{\TBE'}$ and sends $(\pp, \ek, \td)$ to $\A$ as an input.
\item For a decryption query $\ct_{\PKEET} = (\vk, \ct, \ct', \sig)$ from $\A$, $\EventForge$ does not occurs, $\B$ queries $(\vk, \ct)$ to $\mathcal{O}^{\Dec}$ and receives $\pt = \TBEDec(\dk^{*}_{\TBE}, \vk, \ct)$.
Then, $\B$ computes $\pt' \leftarrow \TBEDec'(\dk_{\TBE'}, \allowbreak \vk, \ct')$.
If $\pt' = H(\pt)$, $\B$ returns $\pt$ to $\A$. Otherwise, $\B$ returns $\bot$ to $\A$.

\item For a challenge query $\mathtt{challenge}$ from $\A$, $\B$ samples $\pt^{*}_{0} \xleftarrow{\$} \PtSp^{\PKEET}(\lambda)$ ande sets $ \pt^{*}_{1}  \leftarrow 0$, sends $(\pt^{*}_{0}, \pt^{*}_{1})$ to $\C^{\TBE}$ as a challenge and receives $\ct^{*} \leftarrow \TBEEnc(\ek^{*}_{\TBE}, \allowbreak \vk^{*}, \pt^{*}_{b^{*}})$ where $b^{*}$ is a random bit sampled by $\C^{\TBE}$. 
$\B$ computes $\ct' \leftarrow \TBEEnc'(\ek'_{\TBE}, \allowbreak \vk^{*}, H(\pt^{*}_{0}))$, $\sig \leftarrow \OTSSign(\sk, (\ct, \ct'))$, and returns $\ct_{\PKEET}^{*} \leftarrow (\vk, \ct^{*}, \ct', \allowbreak \sig)$ to $\A$.
\item After receiving the final guess $\pt_{\sfguess}$ from $\A$, if $\pt_{\sfguess} = \pt^{*}_{0}$, $b^{*}_{\sfguess} \leftarrow 0$. Otherwise $b^{*}_{\sfguess} \leftarrow 1$.
$\B$ returns $b^{*}_{\sfguess}$ to $\C^{\TBE}$.
\end{itemize}
In the case of $b^{*} = 0$, $\B$ simulates $\sfGame^{\sfI}_{1, \A}$. 
In the case of $b^{*} = 1$, $\B$ simulates $\sfGame^{\sfI}_{2, \A}$.
From this fact,  we can obtain the following bound.

\begin{equation*}
\begin{split}
&\left| \Pr[b^{*}_{\sfguess} = b^{*}] - \frac{1}{2}  \right| \\
&=  \left| \frac{1}{2}\Pr[b_{\sfguess} = b^{*} | b^{*} = 0] + \frac{1}{2}\Pr[b_{\sfguess} = b^{*} | b^{*} = 1] - \frac{1}{2} \right|  \\
&= \left|  \frac{1}{2} \Pr[b_{\sfguess} = b^{*} | b^{*} = 0]   + \frac{1}{2} \left(1-\Pr[b_{\sfguess} = 0 | b^{*} = 1] \right) - \frac{1}{2} \right|\\
&=\left|  \Pr[\sfGame^{\sfI}_{1, \A}\Rightarrow 1] - \Pr[\sfGame^{\sfI}_{2, \A}\Rightarrow 1] \right|
\end{split}
\end{equation*}
If $\TBE$ satisfies the IND-selTag-wCCA security, $\left| \Pr[b^{*}_{\sfguess} = b^{*}] \right| = \negl(\lambda)$ holds.
Thus, we see that $\left| \Pr[\sfGame^{\sfI}_{2, \A} \Rightarrow 1 ]  - \allowbreak \Pr[\sfGame^{\sfI}_{1, \A} \Rightarrow 1]\right| = \negl(\lambda)$ holds.
\par \hfill (Lemma \ref{OWG1toG2}) \qed
\end{proof*}

\begin{lemma}\label{OWG2toG3}
If $\mathcal{H}$ satisfies the collision resistance property, 
$$|\Pr[\sfGame^{\sfI}_{3, \A} \Rightarrow 1 ]  - \allowbreak \Pr[\sfGame^{\sfI}_{2, \A} \Rightarrow 1]| = \negl(\lambda)$$ holds.
\end{lemma}

\begin{proof*}{Proof of Lemma \ref{OWG2toG3}}
We confirm that Lemma \ref{OWG2toG3} holds.
The difference between $\sfGame^{\sfI}_{2, \A}$ and $\sfGame^{\sfI}_{3, \A}$ is occurs when $\A$ outputs $\pt_{\sfguess}$ which satisfies $\pt_{\sfguess} \neq \pt^{*}$. 
In this case, the collision $(\pt_{\sfguess}, \pt^{*})$ of $\mathcal{H}$ is found.
Let $\EventColl$ be the event that $\A$ outputs $\pt_{\sfguess}$ which satisfies the above.
From the collision resistance property of $\mathcal{H}$, we have $\Pr[\EventColl] = \negl(\lambda)$.
Thus, we have 
\begin{equation*}
|\Pr[\sfGame^{\sfI}_{3, \A} \Rightarrow 1 ]  - \allowbreak \Pr[\sfGame^{\sfI}_{2, \A} \Rightarrow 1]| \leq  \Pr[\EventColl] = \negl(\lambda).
\end{equation*}
\par \hfill (Lemma \ref{OWG2toG3}) \qed
\end{proof*}

\begin{lemma}\label{OWG3}
If $\mathcal{H}$ satisfies the one-wayness, $\Pr[\sfGame^{\sfI}_{3, \A}\Rightarrow 1]   = \negl(\lambda)$ holds.
\end{lemma}

\begin{proof*}{Proof of Lemma \ref{OWG3}}
Lemma \ref{OWG3} is obtained by constructing a reduction $\B$.
We briefly explain how to obtain $\B$. 
$\B$ takes as an input an one-wayness instance $(H^{*}, y^{*})$.
Then, $\B$ sets $H \leftarrow H^{*}$ and simulates $(\pp, \ek)$ and the decryption oracles by following $\sfGame^{\sfI}_{3, \A}$.
For a challenge query $\mathtt{challenge}$ from $\A$, $\B$ sets $\ct \leftarrow \TBEEnc(\ek_{\TBE}, \vk^{*}, 0)$, $\ct' \leftarrow \TBEEnc'(\ek_{\TBE'}, \vk^{*}, y^{*})$ and simulates other element for the challenge cipher $\ct_{\PKEET}^{*}$ by following $\sfGame^{\sfI}_{3, \A}$.
After receiving final output $\pt_{\sfguess}^{*}$ from $\A$, $\B$ outputs $\pt_{\sfguess}^{*}$.
It $\A$ wins the game (i.e., $\sfGame^{\sfI}_{3, \A} \Rightarrow 1 $), $H(\pt^{*}) = y^{*}$ holds.
From these fact, we see that if $\mathcal{H}$ satisfies the one-wayness, $\Pr[\sfGame^{\sfI}_{3, \A} \Rightarrow 1]$ is $\negl(\lambda)$.
\par \hfill (Lemma \ref{OWG3}) \qed
\end{proof*}

From Lemma \ref{OWG0toG1} to \ref{OWG3}, we have
\begin{equation*}
\begin{split}
&\Pr[\sfGame^{\sfOWCCATI}_{\PKEET_{\Ours}, \A}(1^{\lambda}) \Rightarrow 1] =  \Pr[\sfGame^{\sfI}_{0, \A} \Rightarrow 1 ] \\
&\leq \sum_{i \in \{0, \dots , 3\}}  |\Pr[\sfGame^{\sfI}_{i, \A}\Rightarrow 1] - \Pr[\sfGame^{\sfI}_{i+1, \A}\Rightarrow 1] | + \Pr[\sfGame^{\sfI}_{3, \A}\Rightarrow 1] = \negl(\lambda).
\end{split}
\end{equation*}
Thus, we conclude Theorem~\ref{Theorem_OW_CCA_Our}.
\par \hfill (Theorem \ref{Theorem_OW_CCA_Our}) \qed
\end{proof*}

\subsection{IND-CCA Security Analysis}
We prove that $\PKEET_{\Ours}$ satisfies the IND-CCA-T-II security.

\begin{theorem}[IND-CCA Security Type-II Adversary]\label{Theorem_IND_CCA_Ourii}
If $\TBE$ and $\TBE'$ satisfy IND-selTag-wCCA security and $\OTS$ satisfies the sOT-EUF-CMA security, then $\PKEET_{\Ours}$ satisfies the IND-CCA security for type-II adversaries.
\end{theorem}

\begin{proof*}{Proof of Theorem \ref{Theorem_IND_CCA_Ourii}}
Let $\A$ be an adversary for the IND-CCA-T-II security for $\PKEET_{\Ours}$ and $\C^{\PKEET}$ be the challenger of the IND-CCA-T-II security game.
We prove the IND-CCA-T-II security for $\PKEET_{\Ours}$ by the following sequential of games $(\sfGame^{\sfII}_{i, \A})_{i \in \{0,\dots, 3\}}$.

\begin{itemize}
\item{$\sfGame^{\sfII}_{0, \A}:$} 
The original IND-CCA-T-II game $\sfGame^{\sfINDCCATII}_{\PKEET_{\Ours}, \A}(1^{\lambda})$.

\item{$\sfGame^{\sfII}_{1, \A}:$} 
This game is identical to $\sfGame^{\sfII}_{0, \A}$ except that we change the timing of generating of $(\vk^{*}, \sk^{*})$ which is used to generate a challenge ciphertext $\ct_{\PKEET}^{*}$ and add the event $\EventForge$.
At the beginning of the game, $\C^{\PKEET}$ runs $(\vk^{*}, \sk^{*}) \leftarrow \OTSKGen(1^{\lambda})$.
Let $\EventForge$ be the event that $\A$ makes a decryption query that satisfies either of the two following conditions.
This game is identical to $\sfGame^{\sfII}_{0, \A}$ except that we change the timing of generating of $(\vk^{*}, \sk^{*})$. This tuple is used to generate a challenge ciphertext $\ct_{\PKEET}^{*} = (\vk^{*}, \ct^{*}, \ct'^{*}, \sig)$.
Moreover, we  introduce  the event $\EventForge$ in $\sfGame^{\sfII}_{0, \A}$.

At the beginning of the game, $\C^{\PKEET}$ runs $(\vk^{*}, \sk^{*}) \leftarrow \OTSKGen(1^{\lambda})$.
Let $\EventForge$ be the event that $\A$ makes a decryption query that satisfies either of the two following conditions.
\begin{itemize}
\item Before the challenge query, $\A$ makes a decryption query on a ciphertext $\ct_{\PKEET}=(\vk, \ct, \ct', \sig)$ such that $\vk = \vk^{*}$ and $\OTSVerify(\vk^{*}, (\ct, \ct'), \allowbreak \sig) =1$ holds.
\item After the challenge query, $\A$ makes a decryption query on a ciphertext $\ct_{\PKEET}=(\vk, \ct, \ct', \allowbreak \sig)$
such that $\vk = \vk^{*} \land \ct_{\PKEET} \neq \ct_{\PKEET}^{*} \land \OTSVerify(\vk^{*}, \allowbreak(\ct, \allowbreak \ct'),\allowbreak \sig) =1$.
\end{itemize}
If the event $\EventForge$ occurs, $\C^{\PKEET}$ ignores the final output $b^{*}$ by $\A$ and sets $b^{*} \xleftarrow{\$} \{0, 1\}$.

\item{$\sfGame^{\sfII}_{2, \A}:$} 
This game is identical to $\sfGame^{\sfII}_{1, \A}$ except that we change the generation of $\ct$ in the challenge ciphertext $\ct_{\PKEET}^{*}$.
For a challenge query $(\pt_{0}, \pt_{1})$, $\C^{\PKEET}$ sets $\ct \leftarrow \TBEEnc(\ek_{\TBE}, \vk^{*}, 0)$.

\item{$\sfGame^{\sfII}_{3, \A}:$} 
This game is identical to $\sfGame^{\sfII}_{2}$ except that we change the generation of $\ct'$ in the challenge ciphertext $\ct_{\PKEET}^{*}$.
For a challenge query $(\pt_{0}, \pt_{1})$, $\C^{\PKEET}$ sets $\ct' \leftarrow \TBEEnc'(\ek_{\TBE'}, \vk^{*}, 0)$.
\end{itemize}

For $i \in \{0, 1, 2, 3\}$, let $\sfGame^{\sfII}_{i.\A} \Rightarrow 1$ be the event that $\sfGame^{\sfII}_{i, \A}$ outputs $1$.
Then the following facts hold.

\begin{lemma}\label{CCAG0toG1}
If $\OTS$ satisfies the sOT-EUF-CMA security, 
\begin{equation*}
|\Pr[\sfGame^{\sfII}_{1, \A} \Rightarrow 1 ] - \allowbreak \Pr[\sfGame^{\sfII}_{0, \A} \Rightarrow 1] | = \negl(\lambda)
\end{equation*}
holds.
\end{lemma}

Lemma \ref{CCAG0toG1} is proven in a similar way to the proof of Lemma \ref{OWG0toG1}.

\begin{lemma}\label{CCAG1toG2}
If $\TBE$ satisfies the IND-selTag-wCCA security, 
\begin{equation*}
|\Pr[\sfGame^{\sfII}_{2, \A} \Rightarrow 1 ]  - \allowbreak \Pr[\sfGame^{\sfII}_{1, \A} \Rightarrow 1]| = \negl(\lambda)
\end{equation*}
holds.
\end{lemma}

\begin{proof*}{Proof of Lemma \ref{CCAG1toG2}}
Lemma \ref{CCAG1toG2} is obtained by the following reduction algorithm $\B$.
Let $\C^{\TBE}$ the challenger of the IND-selTag-wCCA security game of $\TBE$.
We briefly explain how to construct $\B$ as follows.

\begin{itemize}
\item Given a security parameter, $\B$ runs $(\vk^{*}, \sk^{*}) \leftarrow \OTSKGen(1^{\lambda})$ and sends $\vk^{*}$ to $\C^{\TBE}$. Then, $\B$ receives $\ek^{*}_{\TBE}$ from $\C^{\TBE}$.
\item $\B$ samples $H \xleftarrow{\$} \mathcal{H}$, $\pp \leftarrow H$, $(\ek_{\TBE}, \dk_{\TBE'}) \allowbreak \leftarrow  \nTBESetup'(1^{\lambda})$, $\ek \leftarrow (\ek^{*}_{\TBE}, \ek_{\TBE'})$ and sends $(\pp, \ek)$ to $\A$ as an input.
\item For a decryption query $\ct_{\PKEET} = (\vk, \ct, \ct', \sig)$ from $\A$, $\EventForge$ does not occurs, $\B$ queries $(\vk, \ct)$ and receives $\pt \leftarrow \TBEDec(\dk^{*}_{\TBE}, \vk, \ct)$.
Then, $\B$ computes $\pt' \leftarrow \TBEDec'(\dk'_{\TBE}, \vk, \ct')$.
If $\pt' = H(\pt)$ holds, $\B$ returns $\pt$ to $\A$. Otherwise, $\B$ returns $\bot$ to $\A$.
\item For a challenge query $(\pt_{0}, \pt_{1})$ from $\A$, $\B$ samples $b \xleftarrow{\$}\{0, 1\}$, sends $(\pt^{*}_{0}, \pt^{*}_{1}) \allowbreak = (\pt_{b}, 0)$ to $\C^{\TBE}$ as a challenge and receives $\ct^{*} \leftarrow \TBEEnc(\ek^{*}_{\TBE}, \allowbreak \vk^{*}, \pt^{*}_{b^{*}})$ where $b^{*}$ is a random bit sampled by $\C^{\TBE}$. 
$\B$ computes $\ct' \leftarrow \TBEEnc'(\ek_{\TBE'}, \allowbreak \vk^{*}, \allowbreak H(\pt^{*}_{b}))$, $\sig \leftarrow \OTSSign(\sk, (\ct, \ct'))$, and returns $\ct_{\PKEET}^{*} \leftarrow (\vk, \ct^{*}, \ct', \allowbreak \sig)$ to $\A$.
\item After receiving final guess $b_{\sfguess}$ from $\A$, $\B$ returns $b^{*}_{\sfguess} \leftarrow b_{\sfguess}$ to $\C^{\TBE}$.
\end{itemize}

In the case of $b^{*} = 0$, $\B$ simulates $\sfGame^{\sfII}_{1, \A}$. 
In the case of $b^{*} = 1$, $\B$ simulates $\sfGame^{\sfII}_{2, \A}$.
From this fact, we can obtain the following bound.
\begin{equation*}
\begin{split}
&\left|\Pr[b^{*}_{\sfguess} = b^{*}] - \frac{1}{2} \right|\\
&=  \left| \frac{1}{2}\Pr[b_{\sfguess} = b^{*} | b^{*} = 0] + \frac{1}{2}\Pr[b_{\sfguess} = b^{*} | b^{*} = 1] - \frac{1}{2} \right|\\
&= \frac{1}{2}\left| \left(\Pr[b_{\sfguess} = b^{*} | \sfGame^{\sfII}_{2, \A}] - \frac{1}{2}\right)  + \left(\Pr[b_{\sfguess} = b^{*} | \sfGame^{\sfII}_{2, \A}] - \frac{1}{2}\right) \right|\\
&\geq \frac{1}{2} \left| \Pr \left[\sfGame^{\sfII}_{2, \A} \Rightarrow 1 \right]  - \allowbreak \Pr \left[\sfGame^{\sfII}_{1, \A} \Rightarrow 1 \right]\right| 
\end{split}
\end{equation*}
If $\TBE$ satisfies the IND-selTag-wCCA security, $\left| \Pr[b^{*}_{\sfguess} = b^{*}] - \frac{1}{2} \right| = \negl(\lambda)$ holds.
Thus, we see that $\left| \Pr[\sfGame^{\sfII}_{2, \A} \Rightarrow 1 ]  - \allowbreak \Pr[\sfGame^{\sfII}_{1, \A} \Rightarrow 1]\right| = \negl(\lambda)$ holds.
\par \hfill (Lemma \ref{CCAG1toG2}) \qed
\end{proof*}

\begin{lemma}\label{CCAG2toG3}
If $\TBE'$ satisfies the IND-selTag-wCCA security, $|\Pr[\sfGame^{\sfII}_{3, \A} \Rightarrow 1 ]  - \allowbreak \Pr[\sfGame^{\sfII}_{2, \A} \Rightarrow 1]| = \negl(\lambda)$ holds.
\end{lemma}

The proof of Lemma \ref{CCAG2toG3} is obtained similarly way to the proof of Lemma \ref{CCAG1toG2}. 
\begin{lemma}\label{CCAG3}
$\Pr[\sfGame^{\sfII}_{3, \A} \Rightarrow 1 ] = 0$ holds.
\end{lemma}
\begin{proof*}{Proof of Lemma \ref{CCAG3}}
We confirm that Lemma \ref{CCAG3} holds.
In both case $b=0$ and $b= 1$ of  $\sfGame^{\sfII}_{3, \A}$, the challenge cipher text is $(\ct, \ct') = (\TBEEnc(\ek_{\TBE}, \vk^{*}, 0), \allowbreak \TBEEnc'(\ek_{\TBE}, \allowbreak \vk^{*}, 0)$.
From this fact, we see that $\Pr[\sfGame^{\sfII}_{3, \A} \Rightarrow 1 ] = 0$.
\par \hfill (Lemma \ref{CCAG3}) \qed
\end{proof*}

From Lemma \ref{CCAG0toG1} to \ref{CCAG3}, we have
\begin{equation*}
\begin{split}
&\Pr[\sfGame^{\sfINDCCATII}_{\PKEET_{\Ours}, \A}(1^{\lambda}) \Rightarrow 1] =  \Pr[\sfGame^{\sfII}_{0, \A} \Rightarrow 1 ]\\
&\leq \sum_{i \in \{0, 1, 2\}}  |\Pr[\sfGame^{\sfII}_{i, \A} \Rightarrow 1] - \Pr[\sfGame^{\sfII}_{i+1, \A} \Rightarrow 1] | + \Pr[\sfGame^{\sfII}_{3, \A} \Rightarrow 1] = \negl(\lambda).
\end{split}
\end{equation*}
Thus, we conclude Theorem \ref{Theorem_IND_CCA_Ourii}.
\par \hfill (Theorem \ref{Theorem_IND_CCA_Ourii}) \qed
\end{proof*}

\section*{Acknowledgement.}
A part of this work was supported by JSPS KAKENHI JP24H00071, JST CREST JPMJCR2113, and JST K Program JPMJKP24U2.
We sincerely thank the anonymous IWSEC 2025 reviewers for identifying the flaw in our security proof and for their constructive suggestions to address it.

\appendix

\section{Fundamental Cryptographic Primitives}\label{Sec:Fundamental_Primitives}

\subsection{Hash Functions}
We review fundamental security notions for hash functions. 
\begin{definition}[Collision Resistance]
Let $\mathcal{H}:=\{H_{i \in I}: X \rightarrow Y\}$ be a family of hash functions.
We say that $\mathcal{H}$ satisfies collision resistance if for any PPT adversary $\A$, $\Pr[H(x) = H(x') | H \xleftarrow{\$} \mathcal{H}, (x, x') \leftarrow \A (H)]$ is $\negl(\lambda)$.
\end{definition}

\begin{definition}[One-Wayness]
Let $H: X \rightarrow Y$ be a hash function.
We say that $H$ satisfies one-wayness if for any PPT adversary $\A$, $\Pr[H(x) = H(x') | x  \xleftarrow{\$}X, y \leftarrow H(x), x' \leftarrow \A(H, y)]$ is $\negl(\lambda)$.
\end{definition}

\subsection{One-Time Signature}
We review a definition of a one-time signature scheme and its security notion. 
\begin{definition}[One-Time Signature Scheme]
A (one-time) digital signature scheme $\OTS$ with a message space $\MsgSp(\lambda)$ is a tuple of PPT algorithms $(\nOTSKGen, \allowbreak \nOTSSign, \nOTSVerify)$.
\begin{itemize}
\item $\nOTSKGen(1^{\lambda}):$ A key generation algorithm (probabilistic) takes as an input a security parameter $1^{\lambda}$. 
It returns a verification key $\vk$ and a signing key $\sk$.
\item $\nOTSSign(\sk, \msg):$ A signing algorithm (probabilistic) takes as an input a signing key $\sk$ and a message $\msg$. 
It returns a signature $\sig$.
\item $\nOTSVerify(\vk, \msg, \sig):$ A verification algorithm (deterministic) takes as an input a verification key $\vk$, a message $\msg$, and a signature $\sig$. 
It returns a bit $b \in \{0, 1\}$.
\end{itemize}
\end{definition}

\noindent{\bf Correctness.}
$\OTS$ satisfies the correctness, if for all $\lambda \in \N$, for all $\msg \in \MsgSp(\lambda)$, $(\vk, \sk) \leftarrow \nOTSKGen(1^{\lambda})$, and $\sig \leftarrow \nOTSSign(\sk, \msg)$, then $\nOTSVerify(\vk, \msg, \sig) = 1$ holds.

\begin{definition}[Strong OT-EUF-CMA Security]
Let $\OTS=(\nOTSKGen, \allowbreak \nOTSSign, \allowbreak  \nOTSVerify)$ be a signature scheme and $\A$ be a PPT algorithm.
Strong one-time unforgeability under chosen message attack (sOT-EUF-CMA) security is defined by the following Strong OT-EUF-CMA security game $\sfGame^{\sOTEUFCMA}_{\OTS, \A}(1^{\lambda})$ between a challenger $\C$ and an adversary $\A$.

\begin{itemize}
\item $\C$ runs $(\vk, \sk) \leftarrow \nOTSKGen (1^{\lambda})$, initialize a set $S \leftarrow \{\}$, and sends $\vk$ to $\A$.
\item $\A$ makes only one-time signing query $\msg$ to the signing oracle~$\mathcal{O}^{\nOTSSign}$.
For a signing query on $\msg$ from $\A$, $\mathcal{O}^{\nOTSSign}$ computes $\sig \leftarrow \nOTSSign(\sk, \msg)$, updates $S \leftarrow S \cup \{(\msg, \sig)\}$, and returns $\sig$ to $\A$.
\item $\A$ finally outputs a forgery $(\msg^{*}, \sig^{*})$ to $\C$ where $\msg^{*} \in \MsgSp(\lambda)$.
\item If $\nOTSVerify(\vk, \msg^{*}, \sig^{*}) = 1 \land (\msg^{*}, \sig^{*}) \notin S$ holds, $\A$ wins the game.
\end{itemize}
The advantage of an adversary $\A$ for the sOT-EUF-CMA security game is defined by $\Adv^{\sOTEUFCMA}_{\OTS, \A}(\lambda) \allowbreak := \Pr[\sfGame^{\sOTEUFCMA}_{\OTS, \A}(1^{\lambda}) \Rightarrow 1]$.
$\OTS$ satisfies the sOT-EUF-CMA security if for any PPT adversary $\A$, $\Adv^{\sOTEUFCMA}_{\OTS, \A}(\lambda)$ is $\negl(\lambda)$.
\end{definition}

\section{DLIN-based TBE \cite{Kil06}}\label{Sec:DLIN_TBE}
In this section, we introduce notations for gap groups, recall the DLIN assumption, and review the DLIN-based TBE scheme by Kiltz \cite{Kil06}.

\paragraph{\bf Notations for Groups.}
For a cyclic group $\G$ of a prime order $p$, we define $\G^{*} := \G \backslash \{1_{\G}\}$ where $1_{\G}$ is the identity element of the group $\G$.

\paragraph{\bf Gap Groups.}
First, we introduce gap groups \cite{OP01}.
In these groups, solving the computational Diffie-Hellman (CDH) problem is believed to be hard, despite the existence of an efficient algorithm for the decisional Diffie-Hellman (DDH) problem.

Next, we introduce a gap parameter generator $\GrGen$.
This is a PPT algorithm that takes $1^{\lambda}$ and returns a description of a cyclic group $\G$ of prime order $p$, where $2^{\lambda} < p < 2^{\lambda+1}$ and the description of a Diffie-Hellman oracle $\DDHVerify$.
We call the tuple $(\G, p, \DDHVerify)$ as a description of the gap group.
The oracle $\DDHVerify$ is a PPT algorithm that takes a tuple $(g, g^{x}, g^{y}, g^{z})$ as an input and distinguishes whether the input is the Diffie-Hellman tuple with overwhelming probability.
Formally, we require that for each  $(\G, p, \DDHVerify) \leftarrow \GrGen(1^{\lambda})$ and each $(g, g^{x}, g^{y}, g^{z})$, 
\begin{equation*}
\Pr[\DDHVerify(g, g^{x}, g^{y}, g^{z}) = [xy == z] ] \geq 1 - \negl(\lambda)
\end{equation*}
where $[xy == z] = 1$ if $xy = z \mod p$ and $[xy == z] = 0$ if $xy \neq z \mod p$.

\paragraph{\bf Decision Linear Assumption.}
We review the DLIN assumption \cite{BBS04}.

\begin{definition}[\cite{BBS04,Kil06}]
Let $\GrGen$ be a gap parameter generator.
The decision linear (DLIN) assumption holds for $\GrGen$, if for any PPT adversary $\A$, the following advantage 
\begin{equation*}
\Adv^{\DLIN}_{\GrGen, \A}(\lambda):=
\left|
\Pr
\left[
b^{*} = b 
:
\begin{split}
&(q, \G, \DDHVerify) \leftarrow \GrGen(1^{\lambda}),\\
&g_{1}, g_{2}, z \xleftarrow{\$} \G^{*}, r_{1}, r_{2}, r \xleftarrow{\$} \Z_{q},\\
& b \xleftarrow{\$} \{0, 1\}, w \leftarrow z^{b(r_{1} + r_{2}) + (1-b)r}\\
& b^{*} \leftarrow \A(q, \G, \DDHVerify, g_{1}, g_{2}, z, g^{r_{1}}_{1}, g^{r_{2}}_{2}, w )
\end{split}
\right]
- \frac{1}{2}
\right|
\end{equation*} 
is $\negl(\lambda)$.
\end{definition}

\paragraph{\bf DLIN-Based TBE \cite{Kil06}.}
Now, we are ready to describe the DLIN-based TBE scheme by Kiltz \cite{Kil06}.
Let $\GrGen$ be a gap parameter generator.
The DLIN-based TBE $\TBE_{\DLIN}$ with a tag space $\TagSp = \Z_{p}$ and a message space $\MsgSp = \G$ is given in Fig. \ref{DLIN-based TBE Const}.

\begin{figure}[h]
\centering
\renewcommand{\arraystretch}{1.2}
\begin{tabular}{|l|}
\hline
$\TBESetup (1^{\lambda}):$\\
~~$(\G, p, \DDHVerify) \leftarrow \GrGen(1^{\lambda})$. \\
~~$g_{1} \xleftarrow{\$} \G^{*}, x_{1}, x_{2}, y_{1}, y_{2} \xleftarrow{\$} \Z^{*}_{p}$, choose $g_{2}, z \in \G$ such that $g_{1}^{x_{1}} = g_{2}^{x_{2}} = z$,\\
~~$u_{1} \leftarrow g^{y_{1}}_{1}, u_{2} \leftarrow g^{y_{2}}_{2}$, return $(\ek, \dk) \leftarrow ((\G, p, g_{1}, g_{2}, z, u_{1}, u_{2}), (x_{1}, x_{2}, y_{1}, y_{2}))$.\\

$\TBEEnc(\ek=(\G, p, g_{1}, g_{2}, z, u_{1}, u_{2}), \sftag \in \Z_{p}, \pt \in \G):$\\
~~$r_{1}, r_{2} \xleftarrow{\$} \Z^{*}_{p}$, $C_{1}, \leftarrow g_{1}^{r_{1}}, C_{2}, \leftarrow g_{2}^{r_{2}}, D_{1}, \leftarrow z^{\sftag \cdot r_{1}} \cdot u_{1}^{r_{1}}, D_{2}, \leftarrow z^{\sftag \cdot r_{2}} \cdot u_{2}^{r_{2}}$,\\
~~$K \leftarrow z^{r_{1} + r_{2}}, E \leftarrow M \cdot K$, return $\ct \leftarrow (C_{1}, C_{2}, D_{1}, D_{2}, E)$.\\

$\TBEDec(\dk=(x_{1}, x_{2}, y_{1}, y_{2}), \sftag, \ct=(C_{1}, C_{2}, D_{1}, D_{2}, E)):$\\
~~$s_{1}, s_{2} \xleftarrow{\$} \Z^{*}_{p}$, $K \leftarrow \frac{C^{x_{1} + s_{1} \cdot (\sftag \cdot x_{1} + y_{1})}_{1} \cdot C^{x_{2} + s_{2} \cdot (\sftag \cdot x_{2} + y_{2})}_{2}}{D^{s_{1}}_{1} \cdot D^{s_{2}}_{2}}$\\
~~return $m \leftarrow E \cdot K^{-1}$.\\
\hline
\end{tabular}
\caption{\small The DLIN-based TBE scheme $\TBE_{\DLIN}$ \cite{Kil06}.}
\label{DLIN-based TBE Const}
\end{figure}

\paragraph{\bf Security of DLIN-Based TBE.}
The security of $\TBE_{\DLIN}$ is proven under the DLIN assumption without the ROM.
\begin{lemma}[\cite{Kil06}]
If the DLIN assumption holds for a gap parameter generator $\GrGen$, $\TBE_{\DLIN}$ satisfies the indistinguishability against selective-tag weak chosen-ciphertext attacks (IND-selTag-wCCA) security.
\end{lemma}

By instantiating $\PKEET_{\Ours}[\TBE, \TBE', \OTS, \mathcal{H}]$ (Section \ref{SubSecGeneconPKEET}) with $\TBE = \TBE' = \TBE_{\DLIN}$, we obtain the pairing-free PKEET without the ROM.

\bibliographystyle{abbrvurl}
\bibliography{ref}

\newpage
\setcounter{tocdepth}{2}
\tableofcontents

\end{document}